\documentclass{article}
\usepackage{graphicx}
\usepackage{amsmath}
\usepackage{subfig}
\usepackage[margin=0.5in]{geometry}

\title{Prey Switching with a Linear Preference Trade-Off}

 \author{Sofia H. Piltz \thanks{Oxford Centre for Industrial and Applied Mathematics, Mathematical Institute, University of Oxford, 24-29 St Giles', Oxford, OX1 3LB, UK ({\tt sofia.piltz@linacre.ox.ac.uk}). The work of this author was supported by Osk. Huttunen Foundation and Engineering and Physical Sciences Research Council through the Oxford Life Sciences Interface Doctoral Training Centre.}
 \and Mason A. Porter\thanks{Oxford Centre for Industrial and Applied Mathematics, Mathematical Institute, University of Oxford, 24-29 St Giles', Oxford, OX1 3LB, UK and CABDyN Complexity Centre, University of Oxford, Oxford, OX1 1HP, UK ({\tt porterm@maths.ox.ac.uk})}
 \and Philip K. Maini \thanks{Centre for Mathematical Biology, Mathematical Institute, University of Oxford, 24-29 St Giles', Oxford, OX1 3LB, UK and CABDyN Complexity Centre, University of Oxford, Oxford, OX1 1HP, UK ({\tt maini@maths.ox.ac.uk})}}

\begin{document}

\maketitle

\begin{abstract}
In ecology, \emph{prey switching} refers to a predator's adaptive change of habitat or diet in response to prey abundance.  In this paper, we study piecewise-smooth models of predator-prey interactions with a linear trade-off in a predator's prey preference. We consider optimally foraging predators and derive a model for a 1 predator-2 prey interaction with a tilted switching manifold between the two sides of discontinuous vector fields.  We show that the 1 predator-2 prey system undergoes a novel adding-sliding-like (center to two-part periodic orbit; ``C2PO") bifurcation in which the prey ratio transitions from constant to time-dependent. Further away from the bifurcation point, the period of the oscillating prey ratio period doubles, suggesting a possible cascade to chaos. We compare our model predictions with data and demonstrate that we successfully capture the periodicity in the ratio between {the} predator's preferred and alternative prey types in data on freshwater plankton. Our study suggests that 
it is useful to investigate prey ratio as a possible indicator of how population dynamics can be influenced by ecosystem diversity.

\end{abstract}

\thispagestyle{plain}

\section{Introduction} Ciliates are a type of {\em protist} (eukaryotic single cells with animal-like behavior) that propel using an undulating movement generated by small hair-like protuberances (called {\em cilia}) that cover the cell body. They are also {\em planktonic}\footnote{We have chosen to follow this generally accepted convention instead of {\em planktic}, which is the correct adjective.}, as they are transported primarily by currents. Ciliates occur in aquatic environments and feed on small phytoplankton, so they constitute an important link between the bottom and higher levels of marine and freshwater food webs \cite{TirokGaedke2007}.
 
The seasonal dynamics of plankton predators and prey, in which a spring bloom in phytoplankton is followed by extensive zooplankton predation and nutrient depletion, is well-documented and arises from a combination of abiotic controls (e.g., irradiance and vertical mixing depth) and biotic factors (e.g., competition, grazing, and predation at different levels of a food web) \cite{Sommer1986}. Importantly, ciliates and their algal prey populations also vary at shorter-than-seasonal temporal scales. During years when the spring bloom lasts for several weeks (equivalent to 15 to 30 ciliate generations), phytoplankton and ciliate biomasses exhibit recurring patterns of increases followed by declines \cite{TirokGaedke2007}. In addition, laboratory experiments and environmental observations both suggest that ciliates exhibit negative selection against certain types of prey, because different ciliate species benefit differently depending on the match between their feeding mode and the prey species that are abundant 
in the prey community \cite{MullerSchlegel1999}. Therefore, it has been suggested that the driving forces for the sub-seasonal temporal variability lie in predator-prey interactions between diverse predator and prey plankton communities, particularly during periods of the year in which environmental conditions are relatively settled \cite{TirokGaedke2010}. 

Coexistence of species in a shared environment can arise from ecological {\em trade-offs} \cite{KneitelChase2004}. Phytoplankton have traits that characterize their maximum growth or photosynthesis rate, their size, or their susceptibility to predation; such properties affect how well an individual performs as an organism. Additionally, performance in ecological function is constrained by limited resources. An organism that invests more energy in predation-defense mechanisms cannot invest as much energy in growth or other facets.  Consequently, because of physiological and environmental constraints, species differentiate to occupy niches \cite{KneitelChase2004}. In an environment that changes seasonally, trade-offs give rise to different optimal periods of time for different species \cite{KneitelChase2004}. For example, an increase in temperature can be manifested as a shift in species composition, as species better suited to the new governing conditions become more abundant. Consequently, temporal variation 
of environment and trade-offs result not only in species coexistence amidst limited resources but also in temporal variation exhibited within a community \cite{KneitelChase2004}. Much research has investigated the relationship between diversity and temporal variability in terms of changes in total biomass and/or in species composition within a community \cite{McCann2000}. More recently, 
many studies in community ecology have emphasized traits as a starting point for investigations of co-occurring species \cite{McGilletal2006, LitchmanKlausmeier2008}. 

A standard Lotka-Volterra model for one predator and two prey that allows diversity in the prey community predicts extinction of the prey type that has a smaller capacity to survive \cite{KorobeinikovWake1999}. Therefore, such a framework is inappropriate for investigations of ciliate-phytoplankton dynamics, because several predators and prey coexist, and there is a known preference towards one of the phytoplankton prey \cite{MullerSchlegel1999}. One way to resolve this discrepancy is to examine adaptive predator behavior in response to changes in prey densities. There have been many investigations of \emph{prey switching} \cite{Murdoch1969}, in which predators express preference for more abundant prey. {For example, prey switching has been demonstrated to promote coexistence of competing prey species \cite{Teramoto1979} and to decrease prey competition due to a shared predator \cite{AbramsMatsuda1996}. As a result, prey switching has been suggested as a candidate mechanism for coexistence in communities 
with diverse prey \cite{AbramsMatsuda2003}. 

In smooth differential-equation models for population dynamics, prey switching can be represented using a Holling type-III functional response \cite{Holling1965}, in which a sigmoid function gives the increased degree of predation with increasing densities of a principal prey. In a 1 predator-1 prey system, a Holling type-III functional response corresponds to a situation in which predation is low at low prey densities but saturates quickly at a high value when prey is abundant. Such a functional response was observed in a system of protist predators and their yeast prey \cite{Gauseetal1936}. A low abundance of yeast resulted in increased sedimentation of yeast at the bottom of a glass tube and the walls, so prey were safe from predation, which resulted in a decrease of predator concentration when the prey concentration was below a threshold. Predator preference towards more abundant prey can be examined explicitly by constructing models with multiple prey in which the densities of the different prey are 
system variables. This has been done when switching is considered to be independent of total prey density \cite{Postetal2000} as well as for density-dependent switching \cite{AbramsMatsuda2003}. The latter accounts for the expectation that it is difficult at low prey densities for a predator to distinguish which prey is more abundant \cite{AbramsMatsuda2003}. Using information on which prey type was consumed last, van Leeuwen et al. derived a functional response for a predator switching between two prey types. When one prey density is constant, this functional response is similar to a Holling type-III response. However, it is similar to a Holling type-II functional response when incorporated into a model for population dynamics of 1 predator and 2 prey. More recently, van Leeuwen et al. generalized this method to the case of a predator switching between multiple prey types \cite{vanLeeuwen2013}.

An alternative approach to studying prey switching is to suppose that predators behave as {\em optimal foragers} instead of explicitly incorporating adaptive predator behavior in response to changing prey densities. According to {\em optimal foraging theory}, a predator's choice to switch prey depends on prey abundances and which diet composition maximizes its rate of energy intake \cite{Stephens1986}. Optimal foraging theory is based on constant population densities, and it aims to predict when a predator will always feed on the preferred prey. An optimal forager also feeds on the alternative prey if doing so does not decrease its rate of energy intake \cite{Stephens1986}. K\v{r}ivan \cite{Krivan1996} showed that an alternative prey can be part of an optimal forager's diet (i.e., that the probability that the predator attacks the alternative prey is $p \in [0,1]$). By contrast, the corresponding 1 predator-2 prey system in which the predator is not an optimal forager predicts that both prey exist and that 
the predator becomes extinct. It has also been shown that models based on optimal foragers adaptively feeding on two prey types that differ in their profitability have a larger parameter space in which all species coexist and reduced population density oscillations compared to models with non-adaptive predators \cite{Krivan1996,KrivanSikder1999,BoukalKrivan1999}. Because similar results have been obtained when considering prey growth that is not limited by prey carrying capacities, optimal foraging has been suggested as a possible mechanistic explanation for phenomena such as species coexistence \cite{KrivanEisner2003}. In addition to the choice of whether to include or exclude the alternative prey, similar models of optimal behavior in 1 predator-2 prey systems have been built to study adaptive change of habitat in a two-patch environment \cite{Krivan1997} and adaptive change of activity level \cite{Krivan2007}. 

When using the principle of optimal foraging, a 1 predator-2 prey interaction can be modelled as a discontinuous dynamical system. {\em Piecewise-smooth dynamical systems} are a class of discontinuous dynamical systems \cite{pw-sBook,pws-scholarpedia} in which continuous temporal evolution of state variables alternates with abrupt events that can be caused by friction, collisions, impacts, switching relays, etc. Piecewise-smooth systems arise traditionally in engineering applications, such as an impact oscillator: the dynamical behavior of the oscillator evolves continuously in time above an impact surface, but it is occasionally disrupted because of an impact between the oscillator and a rigid surface \cite{pw-sBook}. Piecewise-smooth dynamical systems have also been used in some applications in biology. For example, in population dynamics, they have been used for the prey switching discussed above as well as to analyze effects on predator-prey population dynamics if a predator population is not allowed to 
be harvested when it is rare \cite{Dercole2007, Kuznetsov2003}. Additionally, piecewise-linear dynamical systems have been used to approximate Hill functions in models of gene regulatory networks \cite{Glass1975, Casey2006} and to describe and quantify synchronization of cattle \cite{Sun2011}.

In the present paper, we apply optimal foraging theory to derive a piecewise-smooth dynamical system that models a predator adaptively feeding on its preferred and an alternative prey. We propose that the predator has two different feeding modes of the preferred prey. The predator behavior is adaptive in the sense that it can choose, depending on the prey densities, to consume the preferred prey either with a large or with a small preference. For simplicity, we assume that consuming a small amount of the preferred prey amounts to not consuming it at all. Consequently, when choosing to consume the preferred prey to a small extent, the predator feeds only on the alternative prey. Similarly, when the predator chooses to feed on the preferred prey with a large preference, the alternative prey is not being predated at all. We take two ecological trade-offs into account. First, we introduce a linear trade-off in the parameter that represents the prey preference. Thus, an increase in the growth in the predator 
population as a result of feeding on the preferred prey comes at a cost of the energy obtained from the alternative prey. Second, to compensate for the preference, we assume that the preferred prey has the advantage of a higher growth rate than that of the alternative prey. We examine the outcome of the population dynamics in such a discontinuous 1 predator-2 prey system as we adjust the steepness of the linear degree of preference trade-off. We thereby discover a previously unknown (at least to our knowledge) bifurcation in piecewise-smooth dynamical systems and provide a possible link between predator-prey dynamics and ecological trade-offs.  

The remainder of this paper is organized as follows. In Section \ref{subsection1}, we derive a piecewise-smooth dynamical system for an optimally foraging predator and two prey in the presence of prey preference trade-off. In Section \ref{subsectionAnalysis}, we investigate this system analytically. We determine the regions in which the system's dynamics evolve along the switching boundary and derive the flow at the switching boundary using Filippov's method. We also present analytical expressions for the equilibrium point at the switching boundary and for the points at which the two vector fields on the different sides of the boundary are tangent to the switching boundary between them. In Section \ref{subsectionNumSimulations}, we investigate the 1 predator-2 prey system numerically. We examine the dynamics as we adjust the slope of the preference trade-off and discover a previously unknown bifurcation in piecewise-smooth systems.  This (``C2PO") bifurcation describes a transition between a center located 
entirely on the switching boundary and a periodic orbit that evolves partly along the boundary and partly outside of it. As the distance to the bifurcation point increases, the periodic orbit experiences a period-doubling, which suggests a possible cascade to chaos. In Section \ref{dataComparison}, we compare simulations of our system with data on planktonic protozoa-algae dynamics. Finally, we discuss and comment on our model assumptions, results, and possible generalizations in Section \ref{Discussion} before concluding in Section \ref{Conclusions}.

\section{Lake Constance Data} \label{dataIntroduction}
Lake Constance is a freshwater lake, with a surface area of 536 km$^2$, that is situated on the Austrian-Swiss-German border. It has been under scientific study for decades, and hourly records of weather conditions---such as temperature, surface irradiance, and wind speed---have been collected since 1979. In addition to such abiotic factors, there are also weekly data for biomass of several phytoplankton and zooplankton species \cite{BaurleGaedke1998}. 

In its natural state, Lake Constance would be categorized as a lake with low level of productivity if there were no excess nutrients from agricultural and other sources of human population in its catchment area. However, the increased sewage water treatment since 1979 has reduced the amount of nutrients that enter Lake Constance, which was categorized as a lake with an intermediate level of productivity (as measured in terms of abundance of nutrients such as nitrogen and phosphorus) when the data were collected \cite{TirokGaedke2010}. Nowadays, Lake Constance has been categorized again as a lake with a low level of productivity. In our comparison between model simulations and data in Section \ref{dataComparison}, we will consider the biomass data for the algal prey of protozoan predators until 1999 that were reported previously in Ref.~\cite{TirokGaedke2006,TirokGaedke2007a}. We concentrate on the early growth season, when the population dynamics are governed predominantly by predator-prey interactions 
rather than by factors such as population dilution due to increased mixing between the upper and lower water masses or strong predation of protists by carnivorous zooplankton population (which develops later in the season) \cite{Sommer1986, TirokGaedke2010}. 

The data that we have received refer to abundance (individuals or cells per ml) and biomass (units of carbon per m$^2$) of a species obtained at least once in a sample of a few ml to a liter of water in Lake Constance between March 1979 and December 1999. When unavailable, the abundance or the biomass have been calculated from each other using species size as a conversion factor. Each sample was collected from a water column with an area of 1 m$^2$ and a depth of 20 m. The full data set includes over 23,000 observations of 205 different phytoplankton species. We have used a subset of these data for the comparison in Section \ref{dataComparison}. In particular, we have selected phytoplankton species that were noted in \cite{TirokGaedke2010} as representative species for preferred and alternative prey of ciliate predators.
 
\section{Piecewise-Smooth 1 Predator-2 Prey Model} 

\subsection{Equations of Motion}\label{subsection1}
We begin the construction of a model for a predator and its preferred and alternative prey 
by assuming a linear trade-off between the preference towards the preferred and alternative preys. In our framework, prey switching occurs because the predator can adjust the extent of consumption of the preferred prey. We assume that there is a trade-off in the prey preference, which effectively is a trade-off in how much energy the predator gains from eating the preferred prey instead of the alternative prey. An increase in specialization towards the preferred prey comes at a cost of the predator population growth obtained from feeding on the alternative prey. For simplicity, we assume that the preference trade-off is linear:
\begin{equation}
	q_2=-a_qq_1+b_q\,,\label{tradeOff}
\end{equation}
where $q_1>0$ is a nondimensional parameter that represents the extent of preference towards the preferred prey, $a_q>0$ is the slope of the preference trade-off, $b_q>0$ is the intercept of the preference trade-off, and $q_2>0$ is the extent of preference towards the alternative prey. Assuming a linear predator mortality and a linear functional response between the predator growth and prey abundance, we define a fitness function that the predator maximizes by the net per capita growth rate
\begin{equation}
	R\equiv \frac{1}{z}\frac{dz}{dt}=eq_1p_1+eq_2p_2-m\,,\label{fitnessFunction}
\end{equation}
where $z$ is the density of the predator population, $p_1$ is the density of the preferred prey, $p_2$ is the density of the alternative prey, $e>0$ is the proportion of predation that goes into predator growth, and $m>0$ is the predator per capita death rate per day. 

We obtain the switching condition that describes when the predator chooses consumption with a large preference towards the preferred prey ($q_1=q_{1_\mathrm{L}}$) or consumption with a small preference towards the preferred prey ($q_1=q_{1_{\mathrm{S}}}$) to maximize fitness by substituting (\ref{tradeOff}) into (\ref{fitnessFunction}) and differentiating $R$ with respect to $q_1$:
\begin{equation}
	\frac{\partial R}{\partial q_1}={ \left(p_1-a_qp_2\right)e}\,.
\end{equation}
Thus, when $\frac{\partial R}{\partial q_1}>0$, the largest feasible $q_{1_{\mathrm{L}}}$ maximizes predator fitness; when $\frac{\partial R}{\partial q_1}<0$, the smallest feasible $q_{1_{\mathrm{S}}}$ maximizes predator fitness. For simplicity, we assume that $q_{1_{\mathrm{S}}}=0$. This implies that the predator switches to the feeding mode of consuming only the alternative prey when predator fitness is maximized by having a small preference for consuming the preferred prey. We also assume (again for simplicity) that prey growth is exponential. This yields the following piecewise-smooth 1 predator-2 prey model: 
\begin{equation}
	\dot{{\mathbf x}}=
	\left[\begin{array}{c} \dot{p_1}\\
	\dot{p_2}\\ 
	\dot{z}\\ \end{array}\right]=
	\left\{ \begin{array}{c}
		f_+=
		\left[\begin{array}{c}
			(r_1-\beta_1z)p_1\\
			r_2p_2\\
			(eq_1\beta_1p_1-m)z
		\end{array}\right]
		\mathrm{, \quad if } \quad h=p_1-a_qp_2>0 \\
		f_-=
		\left[\begin{array}{c}
			 r_1p_1\\
			(r_2-\beta_2z)p_2\\
			(eq_2\beta_2p_2-m)z
		\end{array} \right]
	 \mathrm{, \quad if } \quad h=p_1-a_qp_2<0
	\end{array} \right\}\,,
\label{1pred2prey}
\end{equation}
where $h=p_1-a_qp_2=0$ determines the switching manifold, $r_1>r_2>0$ are the respective per capita growth rates of the preferred and alternative prey, and $\beta_1$ and $\beta_2$ are the respective death rates of the preferred and alternative prey due to predation. In our numerical simulations in Section \ref{subsectionNumSimulations}, we take $\beta_1=\beta_2$. Hence, we assume that the predator exhibits adaptive feeding behavior by adjusting preference rather than attack rate to the governing prey densities. We will specify (\ref{1pred2prey}) at the switching manifold $h=p_1-a_qp_2=0$ in Section \ref{slidingvectorfield}. Production-to-biomass ratio (where the biomass is the mass of all living and dead organic matter and the production represents the increase in the biomass produced by phytoplankton organisms) calculated from measurements in Lake Constance (see Section \ref{dataIntroduction}) can be used as an index for phytoplankton growth. These data suggest that typical values for phytoplankton per 
capita growth rate vary approximately from $0.2$/day to $0.6$/day during the course of a year.

Because of the prey preference, the predator exerts more grazing pressure on the preferred prey than on the alternative prey. Such an advantage for the alternative prey can be explained by a difference in the use of limited nutrients. For example, the alternative prey might invest resources in building defense mechanisms, such as a hard silicate cover, that is difficult for the predator to digest. As a result, the alternative prey has fewer resources left for population growth than the preferred prey (which does not have as good a defense against the predator). To compensate for the difference in preference, we assume that the growth rate of the preferred prey is greater than that of the alternative prey.

\subsection{Analysis of the 1 Predator-2 Prey Model}
\label{subsectionAnalysis}
In this section, we review some definitions from piecewise-smooth dynamical systems and then examine their manifestation in the 1 predator-2 prey system (\ref{1pred2prey}). We will validate our analytical results (and also obtain additional insights) using numerical simulations in Section \ref{subsectionNumSimulations}. 

\subsubsection{Basic Definitions from Piecewise-Smooth Dynamical Systems} \label{subsubsectionBasicDefinitions}
The piecewise-smooth 1 predator-2 prey system with Lotka-Volterra interaction terms in (\ref{1pred2prey}) contains a jump in the derivative of the population densities across the discontinuity boundary $h=0$. This jump makes (\ref{1pred2prey}) a {\em Filippov system}, so its dynamics can evolve along the switching boundary. As shown in Fig.~\ref{dynamicsAtTheSwichingBoundary}, there are three types of such dynamics in piecewise-smooth systems. When the switching boundary is attracting from both sides of the discontinuity, the system is said to exhibit {\em sliding} [panel (a)]. {\em Crossing} occurs when trajectories starting from one side of the discontinuity pass across the switching boundary without following the sliding field on the boundary [panel (b)]. If both of the vector fields point outward from the discontinuity, then the region is defined as a region in which {\em escaping} occurs [panel (c)]. 

\begin{figure}[ht] 
\centering
   \subfloat[sliding]{ 
\includegraphics[width=0.2\textwidth]{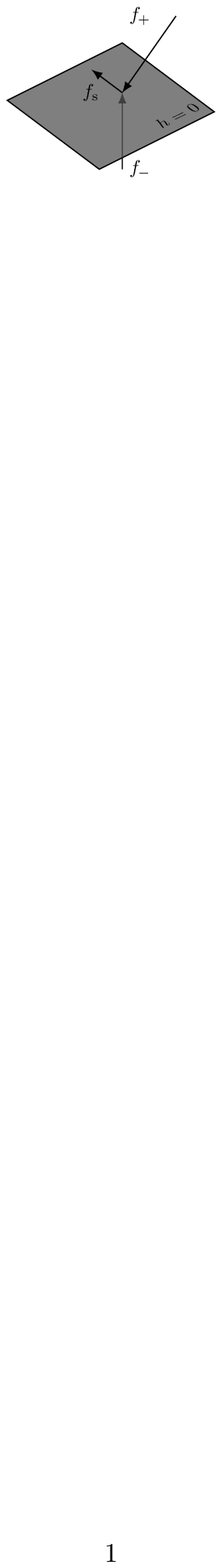}

}    \hspace{0.4in} 
   \subfloat[crossing] {
\includegraphics[width=0.2\textwidth]{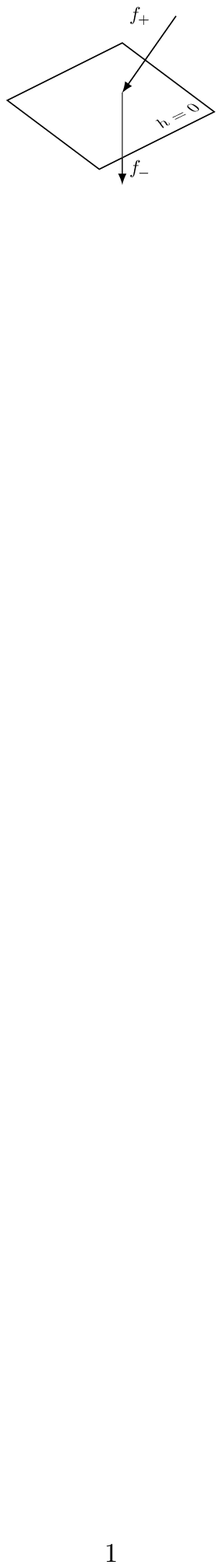}
}  \hspace{0.4in} 
    \subfloat[escaping] {
\includegraphics[width=0.2\textwidth]{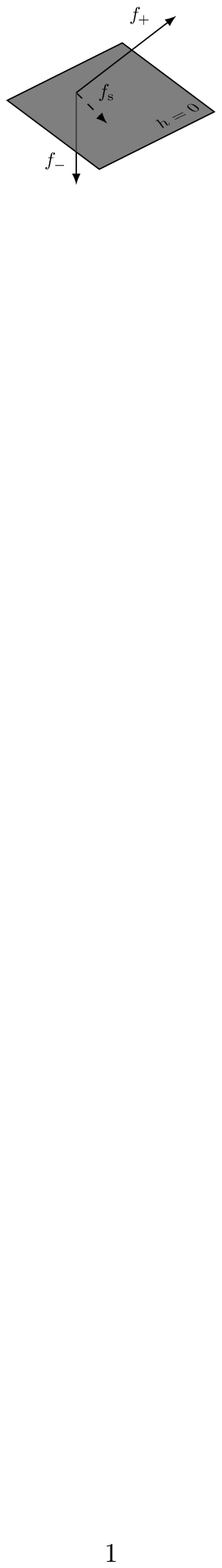}
} 
\caption{The three possible types of dynamics in a piecewise-smooth system close to the switching manifold $h=0$: (a) sliding along the sliding vector field $f_s$, (b) crossing, and (c) escaping (i.e., unstable sliding).} 
\label{dynamicsAtTheSwichingBoundary} 
\end{figure} 

One determines the boundaries of sliding, crossing, and escaping by computing the points at which there is a tangency between the vector fields $f_-$ or $f_+$ and the discontinuity boundary $h=0$ (see Fig.~\ref{tangencies}). Determining the tangencies is crucial for studying the behavior of a system at a switching boundary, and it constitutes the first step for studying how dynamics in a piecewise-smooth dynamical system differ from that in a smooth dynamical system \cite{Colombo2011}. 

There are three basic tangencies between a piecewise-smooth vector field and a switching boundary. In a \emph{fold}, a vector field has a quadratic tangency with a switching boundary [see Fig.~\ref{tangencies} (a)]. In a \emph{cusp}, the tangency is cubic [see Fig.~\ref{tangencies} (b)]. Finally, a {\em two-fold} occurs when two folds intersect, and there is a quadratic tangency between a switching manifold and each side of a vector field [see Fig.~\ref{tangencies} (c)]. In Section \ref{tangencyPoints}, we will determine the points at which one of the vector fields in (\ref{1pred2prey}) has a cubic or quadratic tangency with the switching boundary $h=p_1-a_qp_2=0$. Although a two-fold can arise in three-dimensional piecewise-smooth dynamical systems, it does not occur in (\ref{1pred2prey}) because the crossing and sliding boundaries do not intersect on the switching manifold. We will derive the regions of crossing and sliding in Section \ref{subsectionRegionsofSliding}.

\begin{figure}[ht] 
\centering
   \subfloat[a fold]{ 
\includegraphics[width=0.2\textwidth]{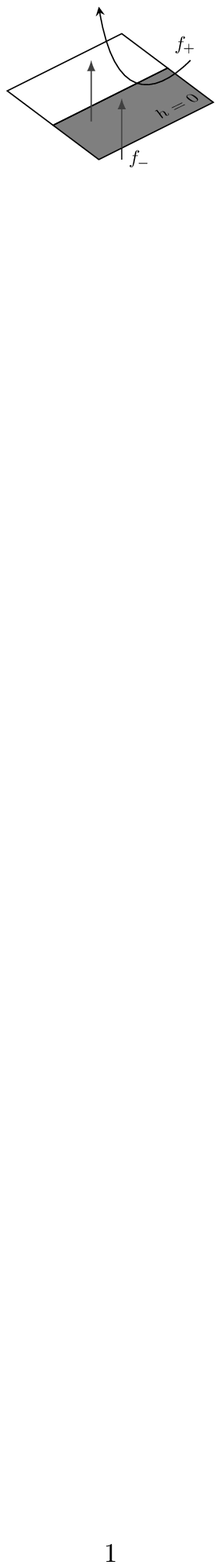}
}    \hspace{0.4in} 
   \subfloat[a cusp] {
\includegraphics[width=0.2\textwidth]{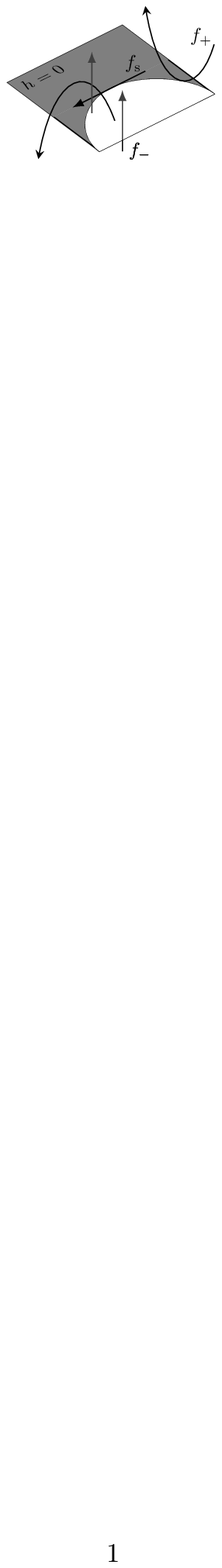}
}  \hspace{0.4in} 
    \subfloat[a two-fold] {
\includegraphics[width=0.2\textwidth]{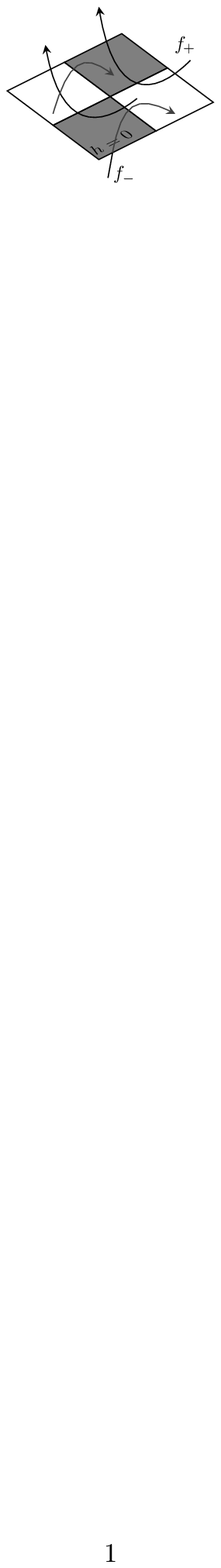}
} 
 \caption{Three basic tangencies between a piecewise smooth vector field and a switching manifold occur at (a) a fold (i.e., a quadratic tangency), (b) a cusp (i.e., a cubic tangency), and (c) a two-fold (i.e., an intersection of two folds). [We have shaded the sliding and escaping regions.]} 
\label{tangencies} 
\end{figure} 

Studying the aforementioned tangencies in a piecewise-smooth system makes it possible to describe bifurcations that can occur when, for example, a limit cycle or an equilibrium point intersects a tangency point on a switching boundary. Such bifurcations are examples of \emph{discontinuity-induced bifurcations} (DIB) \cite{pw-sBook}. In particular, in a system with three or more dimensions, such as the 1 predator-2 prey system (\ref{1pred2prey}), all generic one-parameter sliding bifurcations occur at either a fold, a cusp, or a two-fold. Moreover, because the switching manifold $h = 0$ in (\ref{1pred2prey}) has codimension 1, all of its one-parameter sliding bifurcations can be categorized into 8 different cases (depending on the type of the tangency) \cite{Jeffreyhogan2011}. 

If there is a cusp, a trajectory situated entirely in the sliding region has a tangency with the boundary. That is, the sliding vector field has a quadratic tangency to the switching boundary. Perturbing the bifurcation parameter from the bifurcation point results in the sliding trajectory leaving the switching plane tangentially; because of the cubic tangency with the vector field, it returns to the sliding region. This is called an \emph{adding-sliding} bifurcation because a trajectory that was entirely a sliding trajectory becomes a trajectory that consists of two sliding trajectories with a non-sliding segment of trajectory in between them \cite{pw-sBook}. A new sliding segment is thereby added to the trajectory. 

In the next four subsections, we determine the regions of sliding, crossing, and escaping for (\ref{1pred2prey}); derive the analytical expression for the sliding flow that is the solution to (\ref{1pred2prey}) at the boundary $h=0$; calculate the equilibrium point and associated eigenvalues of the sliding vector field; and compute analytical expressions for points at which there is either a quadratic or a cubic tangency between one of the vector fields and the switching boundary. 

\subsubsection{Regions of Sliding, Crossing, and Escaping}
\label{subsectionRegionsofSliding}
The regions in which (\ref{1pred2prey}) undergoes sliding or escaping are located where the component of $f_+$ normal to $h = 0$ has the opposite sign to the component of $f_-$ normal to $h = 0$. The switching boundary is then either attracting or repelling from both sides of the boundary. The condition for sliding or escaping to occur in (\ref{1pred2prey}) is
\begin{equation}
	\mathcal{L}_{f_+}h\mathcal{L}_{f_-}h=\left(a_qp_2\right)^2\left[(r_1-r_2)^2-z^2\right]<0\,,
\end{equation}
where $\mathcal{L}$ denotes the Lie derivative along the flow $f$ and is defined as $\mathcal{L}_{f_{\pm}}=f_{\pm} \cdot \nabla =\dot{{\mathbf x}}\big|_{f_{\pm}}\cdot \frac{d}{d {\mathbf x}}$. Thus, either sliding or escaping occurs in (\ref{1pred2prey}) for $z^2 > (r_1-r_2)^2$. When $z^2<(r_1-r_2)^2$, the system in (\ref{1pred2prey}) crosses the switching boundary because the components of $f_+$ and $f_-$ normal to $h = 0$ have the same sign. Therefore, trajectories that start from one side of the boundary pass through $h = 0$ without evolving along it. 

In stable sliding regions, the components normal to $h = 0$ of both $f_+$ and $f_-$ point towards the switching manifold and trajectories reach sliding motion in finite time. This occurs in (\ref{1pred2prey}) when 
\begin{equation}
	\mathcal{L}_{f_+}<0<\mathcal{L}_{f_-} \implies -z<r_1-r_2<z\,.
\end{equation}
Escaping (i.e., unstable sliding) occurs because the components normal to $h = 0$ of both $f_+$ and $f_-$ point away from $h = 0$. Trajectories that hit these regions are repelled from the switching manifold in finite time. Escaping motion is unattainable in simulations in forward time. Escaping occurs in (\ref{1pred2prey}) when 
\begin{equation}
	\mathcal{L}_{f_+}>0>\mathcal{L}_{f_-} \implies -z>r_1-r_2>z\,.
\end{equation}

\subsubsection{Sliding Vector Field}
\label{slidingvectorfield}
The solution to (\ref{1pred2prey}) at the discontinuity $h=0$ can be expressed using Filippov's differential inclusion \cite{Filippov1988}. According to Filippov's method, the flow of (\ref{1pred2prey}) at $p_1 = a_qp_2$ is determined by a linear convex combination of the two vector fields $f_-$ and $f_+$ as follows: 
\begin{equation}
	f_s=(1-\alpha)f_-+\alpha(x) f_+\,,
\label{fs}
\end{equation}
where 
\begin{equation}
	{\alpha(x)}=\frac{\mathcal{L}_{f_-}}{\mathcal{L}_{f_-}-\mathcal{L}_{f_+}} \in [0,1]\,.
\label{alphaInFs} 
\end{equation}
If $\alpha=0$, then the sliding flow $f_s$ is given by $f_-$, so $h<0$. Similarly, if $\alpha =1$, then $f_s=f_+$, which implies that $h<0$. Thus, employing (\ref{fs}) and (\ref{alphaInFs}), at $h=0$, we see that the dynamics of (\ref{1pred2prey}) are governed by the sliding vector field 
\begin{equation}
	f_s=\frac{1}{2} 
\left[\begin{array}{c}
	(r_1+r_2-z)p_1\\
	(r_1+r_2-z)p_2\\
	eq_1p_1(r_1-r_2+z)+eq_2p_2(r_2-r_1+z)-2mz
\end{array} \right]\,.
	\label{slidingFlow}
\end{equation}

\subsubsection{The Equilibrium Point}
The sliding vector field (\ref{slidingFlow}) has an equilibrium point when $f_s(\tilde{p}_1, \tilde{p}_2, \tilde{z})=0$. The equilibrium of the sliding flow located at the switching boundary is called a {\em pseudoequilibrium}. For the system (\ref{1pred2prey}), there is a nontrivial pseudoequilibrium point at 
\begin{align}
	\tilde{p}_1  &= \frac{a_qm(r_1+r_2)}{e(q_1a_qr_1+q_2r_2)}\,, \nonumber \\
	\tilde{p}_2  &= \frac{m(r_1+r_2)}{e(q_1a_qr_1+q_2r_2)}\,, \label{pseudoEquilibrium} \\
	\tilde{z} &= r_1+r_2 \nonumber \,.
\end{align}
By evaluating the Jacobian of $f_s$ at $(\tilde{p}_1, \tilde{p}_2, \tilde{z})$, we find that its three eigenvalues are $\lambda_1=0$ and the complex conjugate pair $\lambda_{2,3}$, which satisfy the characteristic equation
\begin{equation}
	\lambda_{2,3} ^2 -\frac{m(r_2-r_1)(q_1a_q-q_2)}{2(q_1a_qr_1+q_2r_2)}\lambda_{2,3}+\frac{m(r_1+r_2)}{2}=0\,.
\end{equation}
The presence of the eigenvalue $\lambda_1 = 0$ suggests that a periodic orbit in (\ref{1pred2prey}) might be born in a novel way.  As we will discuss in detail in Section \ref{addingSlidingSection}, this does indeed turn out to be a previously unknown bifurcation. The eigenvalues $\lambda _{2,3}$ have negative real parts when $a_q>q_2/q_1$, are pure imaginary when $a_q=q_2/q_1$, and have positive real parts when $a_q<q_2/q_1$.

\subsubsection{Tangency Points}\label{tangencyPoints}
At the boundary between sliding and crossing regions, $f_-$ or $f_+$ become tangent to the switching manifold $h=p_1- a_qp_2=0$. At a fold, one vector field has a vanishing first Lie derivative $\mathcal{L}_{f_{\pm}}=0$ (but $\mathcal{L}_{f_\mp} \neq 0$) and a nonvanishing second Lie derivative of $\mathcal{L}^2_{f_{\pm}}=(\mathcal{L}_{f_{\pm}})^2\neq0$. The other vector field satisfies $\mathcal{L}_{f_{\mp}}=0$, and the gradient vectors of $h$ and $\mathcal{L}_{f_{\pm}}$ must be linearly independent of each other \cite{Colombo2011}. The tangency condition for a fold in (\ref{1pred2prey}) is then
\begin{equation}
	\mathcal{L}_{f_-}=\mathcal{L}_{f_+}=0 \implies z=\pm(r_1-r_2)\,.
\end{equation}

At a cusp, a vector field has a cubic tangency to a boundary. This occurs when both $\mathcal{L}_{f_{\pm}}=0$ and $\mathcal{L}^2_{f_{\pm}}=0$. In addition, the sliding vector field has a quadratic tangency to the sliding boundary. Thus, the conditions $\mathcal{L}^3_{f_{\pm}} \neq 0$ and $\mathcal{L}_{f_{\mp}}\neq0$ must hold. Additionally, the gradient vectors of $h$, the first Lie derivative $\mathcal{L}_{f_{\pm}}$, and the second Lie derivative $\mathcal{L}^2_{f_{\pm}}$ are required to be linearly independent \cite{Colombo2011}. For $f_+$ in the 1 predator-2 prey system (\ref{1pred2prey}), the quadratic tangency is given by 
\begin{equation}
	\mathcal{L}^2_{f_{+}}=a_qp_2\left[(r_1-z)^2-r_2^2-z(eq_1a_qp_2-m)\right]\,.
\label{foldForFPlus}
\end{equation}
Substituting $\mathcal{L}_{f_{+}}=0$ into (\ref{foldForFPlus}) yields a cusp at $(p_1,p_2,z)=(m/(eq_1),m/(eq_1a_q),r_1-r_2)$. Similarly, for $f_-$, the condition for a quadratic tangency is 
\begin{equation}
	\mathcal{L}^2_{f_{-}}=a_qp_2\left[(r_2-z)^2-z(m-q_2p_2)+r_1^2\right]\,.
\label{foldForFMinus}
\end{equation}
Substituting $\mathcal{L}_{f_{-}}=0$ into (\ref{foldForFMinus}) yields a cusp at $(p_1,p_2,z)=(a_qm/(eq_2),m/(eq_2),r_2-r_1)$.

\subsection{Numerical Simulations} \label{subsectionNumSimulations}
We now treat the slope of the preference trade-off $a_q$ as a bifurcation parameter and study the 1 predator-2 prey system (\ref{1pred2prey}) numerically. We are interested in the dynamics of (\ref{1pred2prey}) as the complex conjugate pair of eigenvalues of the pseudoequilibrium crosses the imaginary axis. 

The parameter $a_q$ gives the slope of the tilted switching manifold, and it corresponds biologically to the slope of the assumed linear trade-off in the predator's preference for prey. A large $a_q$ corresponds to a situation in which a small increase in the predator's preference towards the preferred prey results in a large decrease in that towards the alternative prey. When $a_q \rightarrow 0$, a small specialization in consuming the preferred prey requires only a small decrease in how much energy the predator gains from eating the alternative prey. At $a_q=0$, the preference trade-off no longer decreases (i.e., it is flat), which corresponds to a situation in which specialization in one prey has no effect on the preference towards consuming the alternative prey.

We simulate (\ref{1pred2prey}) numerically in three different cases: (1) $a_q>q_2/q_1$, (2) $a_q=q_2/q_1$, and (3) $a_q<q_2/q_1$. These correspond, respectively, to (1) $\lambda_1=0$ and the complex conjugate pair $\lambda_{2,3}$ with negative real parts, (2) $\lambda_1=0$ and the complex conjugate pair $\lambda_{2,3}$ with real part $0$, and (3) $\lambda_1=0$ and the complex conjugate pair $\lambda_{2,3}$ with positive real parts. To obtain our numerical solutions, we use the method developed in Ref.~\cite{Piiroinen2008} for solving Filippov systems.  

\subsubsection{The Equilibrium Point}
From our analytical results, we know that the pseudoequilibrium has a $0$ eigenvalue and a complex conjugate pair of eigenvalues with negative real part when $a_q>q_2/q_1$. Accordingly, numerical simulations of (\ref{1pred2prey}) in this parameter regime have trajectories that converge to the pseudoequilibrium in (\ref{pseudoEquilibrium}). See Fig.~\ref{pseudoequilibriumPaperDraft} for an example phase portrait. 

\begin{figure}[ht] 
\begin{center}
\includegraphics[width=0.5\textwidth]{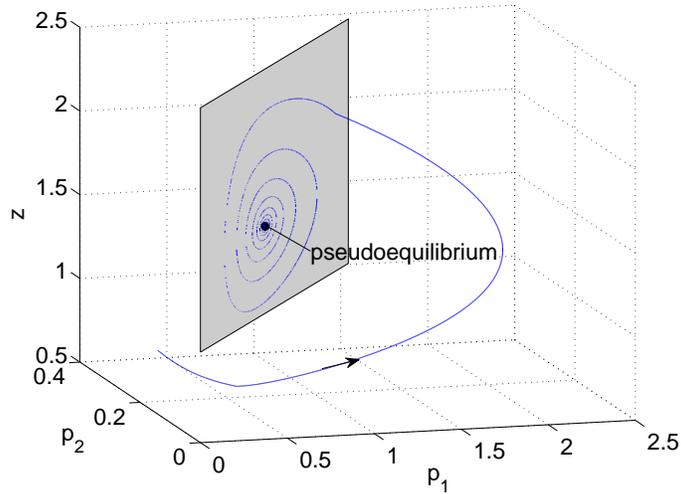}
\caption{Phase portrait for the 1 predator-2 prey model (\ref{1pred2prey}) for parameter values $a_q=4$, $q_1=1$, $q_2=0.5$, $r_1=1.3$, $r_2=0.26$, $e=0.25$, $m=0.14$, and $\beta_1=\beta_2=1$. The system converges to the pseudoequilibrium (black circle) given by (\ref{pseudoEquilibrium}). The predator's diet is composed of both prey types when the system dynamics evolves along the switching boundary $h = 0$ (shaded) according to the sliding vector field (\ref{slidingFlow}).} 
\label{pseudoequilibriumPaperDraft} 
\end{center}
\end{figure} 

\subsubsection{Sliding Centers}
At $a_q=q_2/q_1$, the complex conjugate pair of eigenvalues $\lambda_{2,3}$ have real part $0$. As the sliding vector field is neither attracting nor repelling in the linearized dynamics, every entirely sliding orbit is a periodic orbit that surrounds the pseudoequilibrium (see Fig.~\ref{slidingCentres}). Thus, the amplitude of a periodic orbit depends on the point at which a trajectory intersects the switching surface. In addition to these sliding periodic orbits, there exist periodic orbits that cross the boundary between sliding and crossing regions, but which initially are not entirely sliding, that converge slowly (with dimensional simulation times on the scale of 10$^3$ days and greater}) to an entirely sliding periodic orbit that is tangent to the sliding-crossing boundary (see Fig.~\ref{slidingCentres}). 

\begin{figure}[ht] 
\begin{center}
\includegraphics[width=0.5\textwidth]{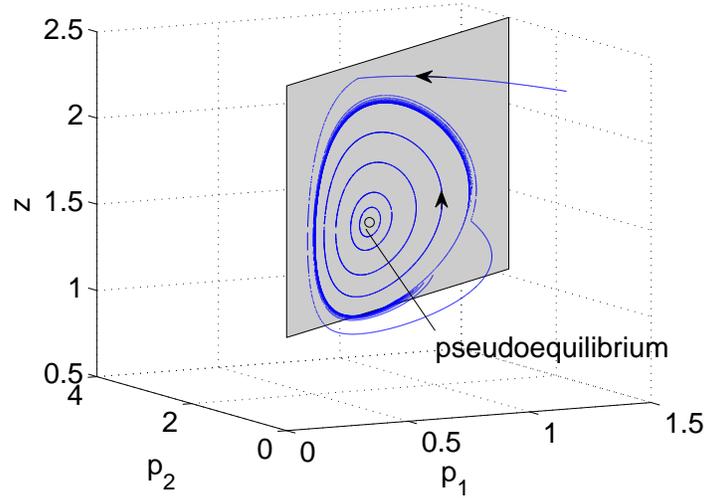}
\caption{Phase portrait for (\ref{1pred2prey}) for parameter values $a_q=0.5$, $q_1=1$, $q_2=0.5$, $r_1=1.3$, $r_2=0.26$, $e=0.25$, $m=0.14$, and $\beta_1=\beta_2=1$. The pseudoequilibrium (black circle) is surrounded by periodic orbits that evolve along the switching manifold $h = 0$ in the stable sliding region (shaded). Periodic orbits that reach the boundary between sliding and crossing eventually converge to the periodic orbit that is tangent to the sliding boundary (blue trajectory).} 
\label{slidingCentres} 
\end{center}
\end{figure} 

\subsubsection{Adding-Sliding Periodic Orbit}
\label{addingSlidingSection}
\begin{figure}[ht] 
\begin{center}
\includegraphics[width=0.5\textwidth]{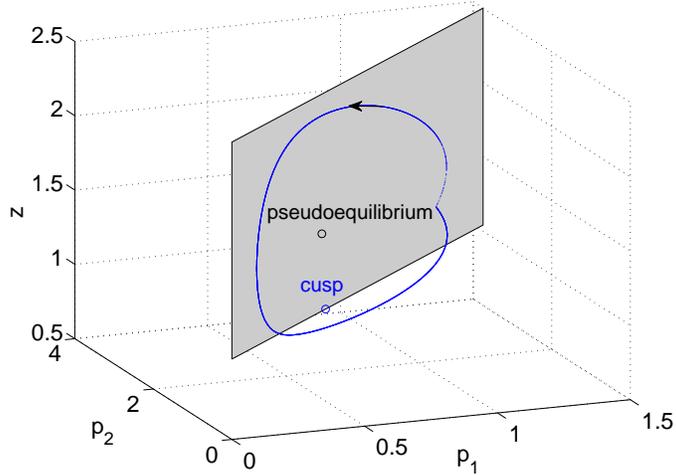} 
\caption{Phase portrait for (\ref{1pred2prey}) for $a_q=0.4$, $q_1=1$, $q_2=0.5$, $r_1=1.3$, $r_2=0.26$, $e=0.25$, $m=0.14$, and $\beta_1=\beta_2=1$. The system has a periodic orbit that leaves the switching manifold (the stable sliding region is shaded) and returns to it because of a cubic tangency between $f_+$ and the switching manifold. Most of the time, the predator's diet is composed of both prey types because the system's dynamics evolves along the switching boundary according to the sliding vector field (\ref{slidingFlow}). With these parameter values, the adding-sliding periodic orbit has a period of approximately 20 days. In a single cycle, the alternative prey is excluded from the diet and $p_1/(a_qp_2)>1$ for approximately 5.5 days.} 
\label{adding_slidingPaperDraft} 
\end{center}
\end{figure}

For $a_q<q_2/q_1$, the pseudoequilibrium (\ref{pseudoEquilibrium}) of the sliding flow is repelling because the complex conjugate pair of eigenvalues $\lambda_{2,3}$ have negative real part. There is also a periodic orbit. From our analytical calculations, we know that the vector field $f_+$ has a cubic tangency with the switching boundary at the cusp at $(p_1,p_2,z)=(m/(eq_1),m/(eq_1a_q),r_1-r_2)$. Because of the cusp, the local flow near the tangency forces the trajectory of the periodic orbit to first leave the switching boundary tangentially and then to return to it. In doing this, the periodic orbit acquires a non-sliding segment before returning to the switching manifold $h = 0$. See Fig.~\ref{adding_slidingPaperDraft} for an example phase portrait.     

\begin{figure}[ht] 
\begin{center}
\includegraphics[width=0.7\textwidth]{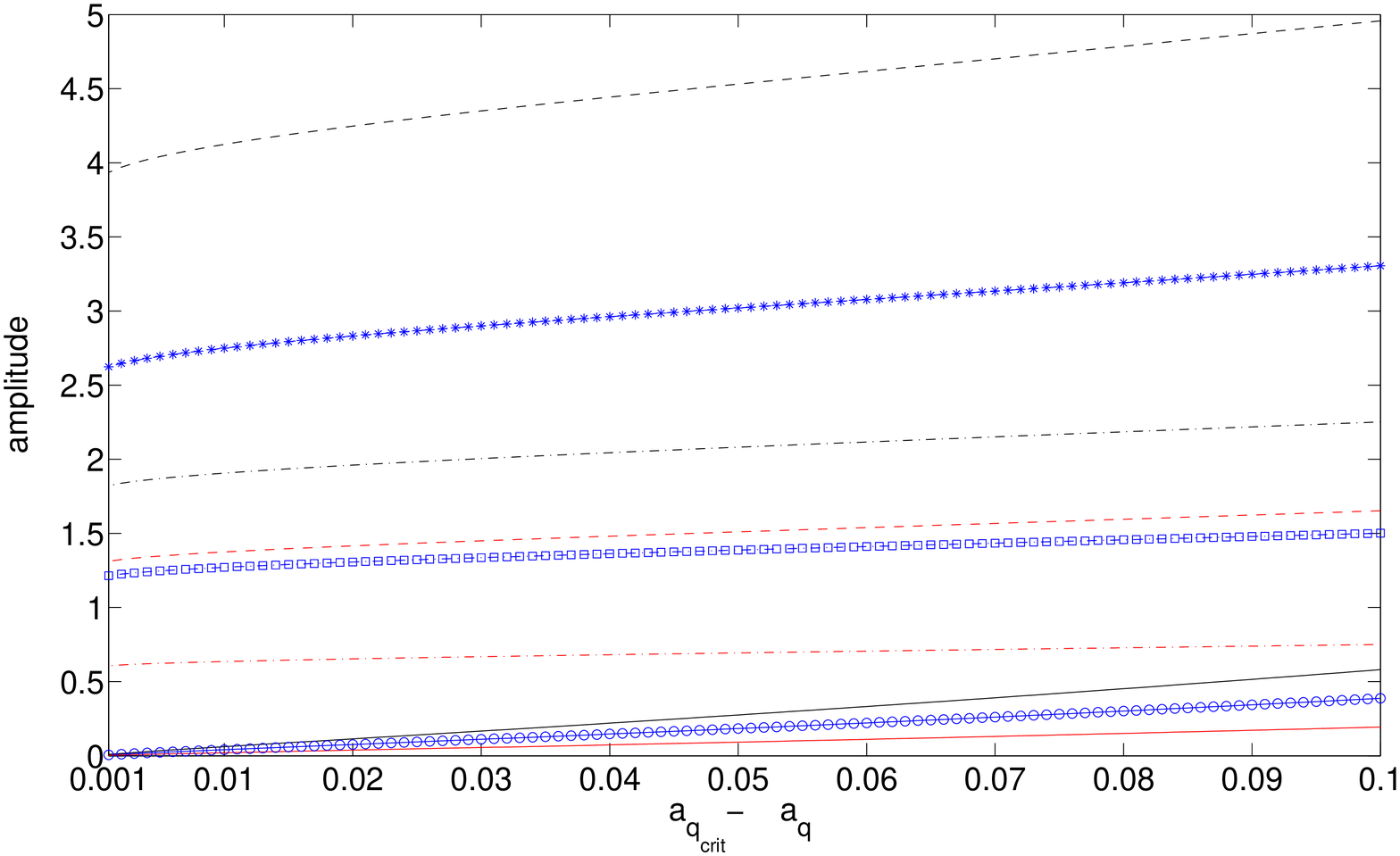} 
\caption{Amplitude of $H-\tilde{H}$ (circles), $G-\tilde{G}$ (squares), and $z-\tilde{z}$ (asterisks) of (\ref{1pred2prey}) [for parameter values $q_1=1$, $q_2=0.5$, $r_1=1.3$, $r_2=0.26$, $e=0.25$, $m=0.14$, and $\beta_1=\beta_2=1$] versus the distance $a_{q{\mathrm{crit}}}-a_q<0.1$ from the bifurcation point $a_{q{\mathrm{crit}}}=q_2/q_1$. We illustrate the possible scaling relations for $a(H-\tilde{H})$ (solid curves), $a(G-\tilde{G})$ (dashed curves), and $a(z-\tilde{z})$ (dash-dotted curves) for three different values: $a=1$ (blue), $a=0.5$ (red), and $a=1.5$ (black).} 
\label{amplitudeScaling} 
\end{center}
\end{figure}

Adding-sliding periodic orbits can be detected using the sliding condition, as there are two separate pieces of sliding trajectories when $p_1/(a_qp_2)=1$. Between the sliding pieces, there is a segment of non-sliding trajectory when $p_1/(a_qp_2)>1$. Using numerical simulations of (\ref{1pred2prey}), we examine how the amplitude of the adding-sliding periodic orbit measured from the pseudoequilibrium $(\tilde{p}_1, \tilde{p}_2, \tilde{z})$ in (\ref{pseudoEquilibrium}) scales with the distance to the bifurcation point $a_{q{\mathrm{crit}}} = q_2/q_1$. For $a_{q{\mathrm{crit}}}-a_q<0.1$, we record the amplitude as the difference in the maximum and minimum values of $H-\tilde{H}$, $G-\tilde{G}$, and $z-\tilde{z}$, where $H=p_1-a_qp_2$, $\tilde{H}=\tilde{p_1}-a_q\tilde{p_2}$, $G=a_qp_1+p_2$, and $\tilde{G}=a_q\tilde{p_1}+\tilde{p_2}$. Based on visual inspection, the scaling near $a_{q{\mathrm{crit}}}$ appears to be linear (see Fig.~\ref{amplitudeScaling}), and we note for comparison that a linear scaling 
relation is known to arise for the  ``generalized Hopf bifurcation" for piecewise-smooth dynamical systems discussed in Ref.~\cite{SimpsonMeiss2007}. In this context, such a generalized Hopf bifurcation refers to a periodic orbit that is born when an equilibrium point of a planar, piecewise-smooth, continuous system crosses the switching boundary. Here, by a ``continuous system," we mean a piecewise-smooth system in which the trajectories always cross the switching boundary without evolving along it. In that generalized Hopf bifurcation, a complex conjugate pair of eigenvalues of a piecewise linear system (obtained by linearizing a piecewise smooth system) has a negative real part on one side of the switching boundary, $0$ real part on the switching boundary, and a positive real part on the other side of the switching boundary \cite{SimpsonMeiss2007}.

The above bifurcation in (\ref{1pred2prey}), in which an adding-sliding periodic orbit arises from a center, has not previously been studied (to our knowledge). Because a center transitions to such a two-part periodic orbit, we call this a \emph{center to two-part periodic orbit (``C2PO") bifurcation}.  In the C2PO bifurcation, adding-sliding periodic orbits are born via a two-event sequence. First, there is a bifurcation reminiscent of the standard Hopf bifurcation from smooth dynamical systems. This arises via the behavior of the eigenvalues of the sliding vector field $f_s$, as the pseudoequilibrium changes from attracting to repelling and a periodic orbit appears. Second, as the distance to the bifurcation point increases, the periodic orbit grows and an adding-sliding bifurcation ensues \cite{pw-sBook}.  

\subsubsection{Period Doubling}
\begin{figure}[ht] 
\begin{center}
\includegraphics[width=0.7\textwidth]{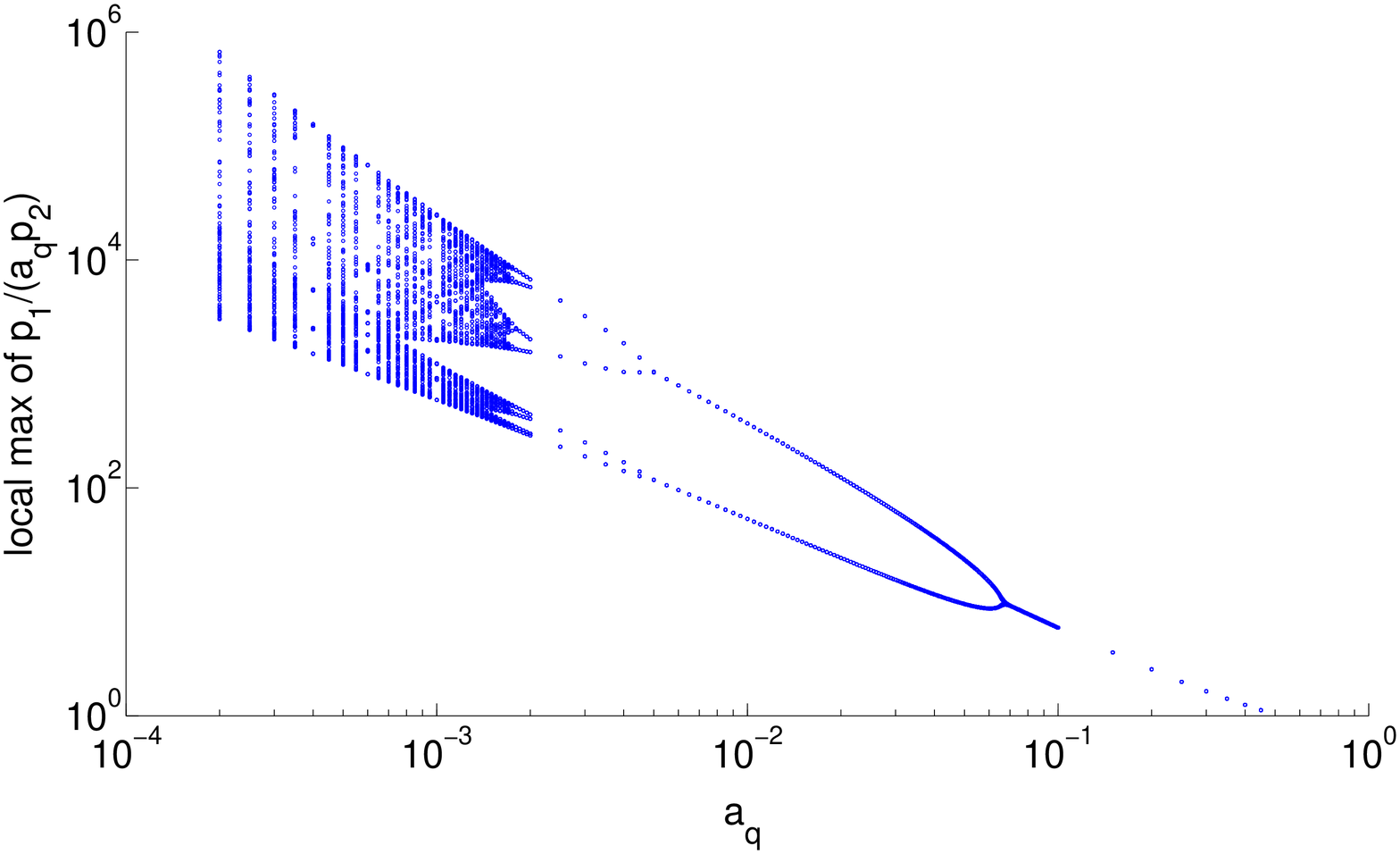} 
\caption{Local maxima in $p_1/(a_qp_2)>1$ as $a_q\rightarrow0$ for (\ref{1pred2prey}) with parameter values $r_1=1.3$, $r_2=0.26$, $e=0.25$, $m=0.14$, {and $\beta_1=\beta_2=1$}.} 
\label{bifurcationDiagram} 
\end{center}
\end{figure}

We show a bifurcation diagram for (\ref{1pred2prey}) by detecting the local maxima of the quantity $p_1/(a_qp_2)>1$ when $a_q \rightarrow 0$ and $b_q \rightarrow q_2$. The period-1 adding-sliding periodic orbit that emerges when $a_q<q_2/q_1$ period-doubles as we decrease $a_q$ away from the bifurcation point. As we illustrate in Fig.~\ref{bifurcationDiagram}, this suggests that there is a cascade to chaos as $a_q \rightarrow 0$. From a biological perspective, $a_q \rightarrow 0$ corresponds to the situation in which there is little decrease in the preference towards the alternative prey if the predator has an increase in specialization towards the preferred prey.

\subsection{Comparison of 1 Predator-2 Prey Model Simulations and Planktonic Protozoa-Algae Data} \label{dataComparison}
In this section, we compare the prey ratio of an adding-sliding period-1 orbit from the 1 predator-2 prey model (\ref{1pred2prey}) with data on cryptophyte and diatom prey collected from Lake Constance in spring.

\begin{figure}[ht] 
\begin{center}
\includegraphics[width=0.7\textwidth]{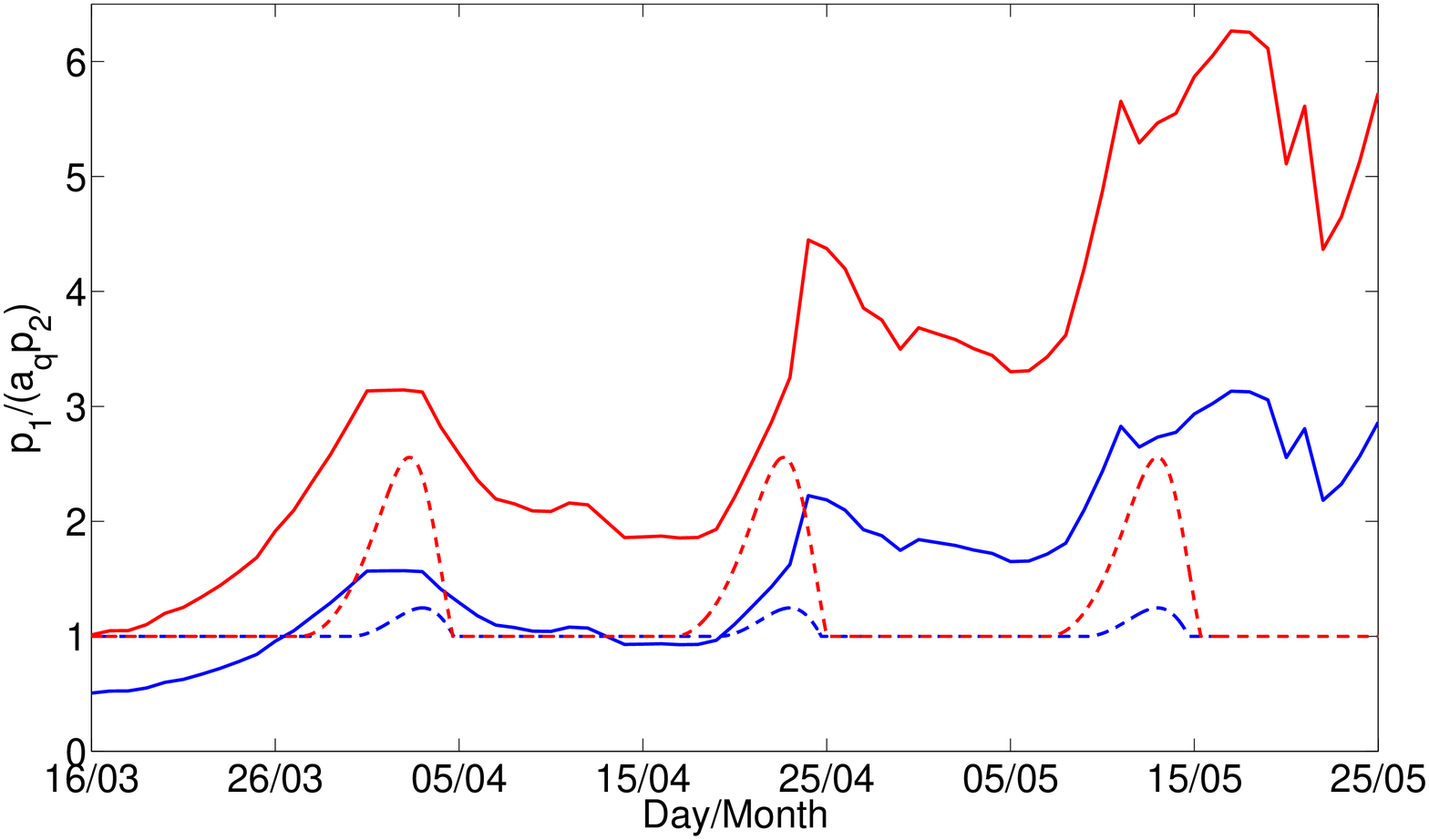} 
\caption{Scaled prey ratio $p_1/(a_qp_2)$ 
for simulations of (\ref{1pred2prey}) (dashed curves) for $a_q=0.4$ (blue), $a_q=0.2$ (red), $r_1=1.3$, $r_2=0.26$, $e=0.25$, $m=0.14$, {and $\beta_1=\beta_2=1$} and the mean data (solid curves) over 1979--1999 for the prey ratio $p_1/(a_qp_2)$ collected in spring in Lake Constance. The preferred prey $p_1$ is composed of data for {\em Cryptomonas ovata}, {\em Cryptomonas marssonii}, {\em Cryptomonas reflexa}, {\em Cryptomonas erosa}, {\em Rhodomonas lens}, and {\em Rhodomonas minuta}. The alternative prey group $p_2$ is composed of data for small and medium-size {\em Chlamydomonas} spp. and {\em Stephanodiscus parvus}. These data, previously reported in Ref.~\cite{TirokGaedke2006,TirokGaedke2007a} were kindly given to us by Ursula Gaedle.} 
\label{simulationAndData} 
\end{center}
\end{figure}

Previous studies of the Lake Constance data set suggest that ciliates are the most abundant herbivorous zooplankton group in spring, whereas cryptophytes and diatoms are the dominant species groups in the phytoplankton community \cite{TirokGaedke2007}. We use 6 different cryptophyte species as the group of preferred prey and 3 different diatoms species to represent ciliates' alternative prey. We include data for phytoplankton whose cell size is sufficiently small in comparison to the size of ciliate predators and which dominate the algal community in Lake Constance in spring \cite{TirokGaedke2007}. For both species groups, we use linear interpolation to obtain intermediate biomass values for each day of the year from approximately bi-weekly measurements. We then calculate the mean of the 20 interpolated yearly data between 1979 and 1999. In Fig.~\ref{simulationAndData}, we compare these data with the prey ratio that we obtain from simulations of (\ref{1pred2prey}). Although our model does not capture the 
increasing trend, it successfully reproduces the periodic pattern early in the growing season when predator-prey interaction can be argued to govern the protist-algae dynamics more than physical driving forces in water masses that are rich in nutrients \cite{Sommeretal2012}.

\section{Discussion}\label{Discussion}
We have combined adaptive predator behavior and ecological trade-offs to model the dynamics of a predator feeding on a preferred and alternative prey as a piecewise-smooth dynamical system. Our model describes an optimal forager that can adaptively change its diet depending on the abundances of its preferred and alternative prey. We assumed a linear trade-off in prey preference and analyzed the dynamics of the system as we adjusted the slope of the trade-off. To compensate for the preference, we assumed that the preferred prey has a larger growth rate than the alternative prey.

Our model predicts a steady-state for the predator and prey populations as long as the trade-off in prey preference is sufficiently steep. In other words, a steady state arises when a small increase in specialization towards the preferred prey would result in a large decrease in how much energy the predator could gain from the alternative prey. As we decrease the slope of the preference trade-off, a periodic orbit appears and period-doubles as the slope approaches $0$. After the bifurcation, the population densities oscillate and the prey ratio is no longer constant. From a biological perspective, a mild trade-off suggests that a predator with a small increase in the energy gained from the preferred prey would experience only a small decrease in the energy gained from the alternative prey. We considered a linear preference trade-off as a first step in studying the effect of such a trade-off in a predator-prey interaction in which prey switching occurs. Although several studies and observations support the 
existence of trade-offs \cite{KneitelChase2004, LitchmanKlausmeier2008}, it is not clear what kind of functional form they take. A previous model for population dynamics and prey switching studied a convex (i.e., concave down) trade-off between the attack rate and the degree of specialization exhibited by a predator \cite{AbramsMatsuda2003}. Concave relationships have been formulated for the consumption of two prey species and the energy obtained from two prey species in a graphical approach for solving the problem of optimal diet \cite{Rapport1971}. To develop a more complete picture of the effects of a preference trade-off on populations dynamics, it would be useful to consider generalizations of our model with nonlinear trade-off functions. 

In prey switching, a predator can adopt either a large or a small consumption of the preferred prey type according to prey abundances. For simplicity, we assumed that the smallest feasible consumption is $0$. Traditionally, prey-switching models based on optimal foraging theory assume that a predator either includes or excludes the alternative prey and that the preferred prey is always part of the diet \cite{Stephens1986}. Despite our simplifying assumption, our model simulations always converge to an attractor in which the preferred prey is not excluded from the diet. We have also constructed an alternative model (in addition to that on which this paper focuses) to describe a predator that eats the preferred prey until it becomes rare, at which point it also eats the alternative prey. In the presence of a linear prey preference, this piecewise-smooth dynamical system has a pseudoequilibrium with a $0$ eigenvalue and a complex conjugate pair of eigenvalues that never crosses the imaginary plane as long as 
the growth rate of the preferred prey is assumed to be larger than that of the alternative prey. Our numerical simulations of this model suggest that such a 1 predator-2 prey system in which the preferred prey is always part of the diet does not exhibit the bifurcation that we see in the 1 predator-2 prey system (\ref{1pred2prey}) in which the predator eats only one prey type at a time.

We have also constructed a smooth analog of (\ref{1pred2prey}) using hyperbolic tangent functions. The two systems produce similar behavior as long as the slope of the hyperbolic tangent is sufficiently large. In particular, both dynamical systems settle to a steady state for $a_q>\frac{q_2}{q_1}$ and to a periodic orbit for $a_q<\frac{q_2}{q_1}$. However, when the slope of the hyperbolic tangent is smaller, the smooth system predicts that the predator and the two prey also coexist at steady-state levels (instead of oscillating) when $a_q<\frac{q_2}{q_1}$. We presented the analytical expression for the pseudoequilibrium (\ref{pseudoEquilibrium}) of the piecewise-smooth system (\ref{1pred2prey}), and we calculated the equilibrium point for the smooth system numerically.

In the present paper, we have interpreted adaptation as flexible feeding behavior of a predator that has two feeding modes for consuming its preferred prey, and we have assumed that it can switch between them in response to prey densities. Recent studies have shown that rapid adaptation of traits affects ecological interactions and can be observed especially in organisms with short lifespans, such as species in plankton communities \cite{Becksetal2010,Jonesetal2009}. Interactions between evolutionary adaptation and population dynamics have been studied using both a framework of \emph{fast-slow} and {\emph{slow-fast dynamical systems}. The former approach has given insight into how evolution of traits is seen in the population dynamics by studying the limit in which trait evolution occurs on a much faster time scale than predator-prey interaction \cite{CortezEllner2010}. In the latter framework, the consequences that ecological dynamics can have on the evolution of traits have been studied with a piecewise-
smooth dynamical system; this assumes that evolution occurs at a much longer time scale than predator-prey interactions \cite{Dercoleetal2006}. Formulating a fast-slow dynamical-system analog of (\ref{1pred2prey}) would enable an interesting comparison of piecewise-smooth and fast-slow dynamical systems and should be insightful for modelling interactions between evolutionary adaptation and population dynamics.

We chose to use exponential prey growth to simplify analytical calculations. An important generalization is to consider logistic prey growth, as nutrient limitation is one of the most important nonliving factors that drive the temporal pattern of phytoplankton growth \cite{Sommer1986}. However, it has been suggested that the importance of nutrient limitation is less pronounced than protist grazing at the beginning of a growth season \cite{Sommeretal2012} for water masses that are rich in nutrients. Although considering logistic prey growth increases the number of parameters in a model (in addition to making analytical calculations significantly more challenging), it has two key advantages: (1) it would expand the suitable time window for comparing simulations with data in water masses that are rich in nutrients, and (2) its less restrictive assumptions yield a 
model that is also reasonable in principle for water masses with low nutrient levels \cite{Sommeretal2012}.

Our 1 predator-2 prey system (\ref{1pred2prey}) reproduces the periodicity but does not capture the increasing trend in the scaled prey ratio during the early growing season in Lake Constance. We speculate that a generalization that allows both the prey growth rates and the preference trade-off slope to be time-dependent might make it possible to also capture the increasing trend. We motivate the time-dependent prey growth rate by the fact that the phytoplankton production-to-biomass ratio (results not shown) calculated for Lake Constance exhibits a linear increase in the prey growth in spring and a decrease in the autumn \cite{Ursulaperscomm}. To motivate the time-dependent slope of the preference trade-off, we remark that there is some evidence for variation of the shape of an ecological trade-off in bacteria in different environmental conditions \cite{JessupBohannan2008}.

As a first step towards studying models of adaptive predator behavior and ecological trade-offs, we started from the simplest case (i.e., a system with 1 predator and 2 prey). In the model (\ref{1pred2prey}) that we have introduced, functional diversity is present only in the prey community and it arises as the difference in prey growth rates. Accordingly, we have chosen species from a large data set to consider representative prey groups and have left prey competition and predator diversity for future work. Ciliates are known to have different modes of predator behavior, and they can be categorized roughly in terms of being more or less selective \cite{Verity1991}. To illustrate a predator that is more selective, we note that some ciliate species hunt as interception feeders that scavenge on food particles and intercept them directly. By contrast, ciliate filter feeders sieve suspended food particles and provide an example of less selective predators. Such diversity in the predator community can be 
represented using different preference trade-offs. This could be studied using a a piecewise-smooth dynamical system with a dimension higher than 3 (or using a fast-slow dynamical system) and which could have more than one switching manifold. In addition, the switching manifolds might intersect with each other. Such a generalization would thus be very interesting (and challenging) to study from both biological and mathematical perspectives. We have already motivated the former. From the latter perspective, we remark that there does not yet exist a general treatment for bifurcations that arise from intersections of switching manifolds when the ambient space has more than 3 dimensions \cite{Colombo2011}.

Our model exhibits a novel (``C2PO") bifurcation, in which the dynamics transitions at the bifurcation point between convergence to a pseudoequilibrium and periodic adding-sliding oscillations through a center. This emergence of adding-sliding orbits differs from the usual way that they emerge in piecewise-smooth dynamical systems \cite{pw-sBook}. The standard mechanism entails the birth of an adding-sliding periodic orbit following a bifurcation in which the eigenvalues of the pseudoequilibrium cross the imaginary plane, so that the pseudoequilibrium changes from attracting to repelling and an entirely sliding periodic orbit is born. As the amplitude of the periodic orbit (which lies entirely on the switching boundary) grows, the sliding periodic orbit eventually becomes tangent to the boundary between sliding and crossing. Finally, this periodic orbit becomes a trajectory that has two sliding segments separated by a non-sliding segment \cite{pw-sBook}. We observed numerically that the amplitude scaling of 
the adding-sliding periodic orbit emerging from the C2PO bifurcation appears to be linear in the distance from the bifurcation, which is also the case for the generalized Hopf bifurcation in piecewise-smooth dynamical systems in Ref.~\cite{SimpsonMeiss2007}. 

\section{Conclusions}\label{Conclusions}
We combined two ecological concepts---prey switching and trade-offs---and used the framework of piecewise-smooth dynamical systems to develop a model of one predator feeding on a preferred and alternative prey. We derived analytical expressions for the pseudoequilibrium, its eigenvalues, and the points for tangencies between the two vector fields and the switching boundary. We confirmed our analytical results with numerical simulations, and we discovered a novel bifurcation in which an adding-sliding periodic orbit is born from a center. Based on numerical simulations close to the bifurcation point, the amplitude of the adding-sliding periodic orbit seems to scale linearly with the distance from the bifurcation point. 

Our model introduces a way to link trade-offs with adaptive predator behavior. We compared the results of our simulations with data on freshwater plankton, but we remark that similar prey-switching models can also be formulated for any other 1 predator-2 prey interaction in which it is viable to use models based on low-dimensional differential equations (i.e., large population size, well-mixed environment, and the use of community-integrated parameters). We also discussed several biologically motivated generalizations of our model. We believe that our current model and these generalizations provide a promising direction for examining possible mechanisms for ecological trade-offs in population dynamics. 

Although we have motivated our investigation primarily from a biological perspective, it is also important to stress the utility of our model for development of new theoretical understanding in piecewise-smooth dynamical systems. The biological and mathematical motivations complement each other very well, and investigating the condition for sliding corresponds to studying a scaled prey ratio, and this in turn offers a possible link between ecological trade-offs and population dynamics. We believe that our model offers an encouraging example of how combining theoretical and practical perspectives can give new insights both for the development of theory of piecewise-smooth dynamical systems as well as for the development of models of population dynamics with predictive power.

\section*{Acknowledgements}
We thank Jorn Bruggeman, Alan Champneys, Stephen Ellner, John Guckenheimer, and John Hogan for helpful discussions. We also thank Ursula Gaedke for useful discussions and sending us data, and we thank Chris Klausmeier and Mike Jeffrey for numerous helpful discussions and comments. The Lake Constance data were obtained within the Collaborative Programme SFB 248 funded by the German Science Foundation.

\end{document}